\documentclass[newstyle,twocolumn,proceedings]{rmaa}

\def\plotBTD#1#2{%
  \expandafter\ifx\csname epsfbox\endcsname\relax
    \immediate\write16{%
        You need to input epsf; I'll do it for you
    }%
    \input epsf
  \fi
  \epsfysize=#2
     \openin 1 #1 \ifeof 1
        \immediate\write16{Can't open #1}%
        \vskip \the\epsfysize
      \else
         \closein 1
         \centerline{\epsfbox{#1}}%
      \fi
}

\title{Compressible MHD Turbulence: Mode Coupling, Anisotropies and Scalings}
\SetYear{2002}
\SetConfTitle{Winds, Bubbles and Explosions: a Conference to Honor John Dyson}

\author{Jungyeon Cho and A. Lazarian 
      \affil{Dept. of Astronomy, University of Wisconsin, USA}}

\shortauthor{Cho and Lazarian}
\shorttitle{MHD Turbulence}

\fulladdresses{\item Dept. of Astronomy, University of Wisconsin,
                475 N. Charter St., Madison, WI53706, USA;\\
                cho@astro.wisc.edu,
                lazarian@astro.wisc.edu. }

\listofauthors{Jungyeon Cho \& A. Lazarian}
\indexauthor{Cho, J.}
\indexauthor{Lazarian, A.}

\SetVolume{00} \SetFirstPage{1} \SetYear{2002}
\ReceivedDate{} 
\AcceptedDate{} 
\resumen{}
\abstract{Compressible turbulence, especially the magnetized version of
it, traditionally has a bad reputation with researchers. 
However, recent
progress in theoretical understanding of incompressible MHD as well as
that in computational capabilities
enabled researchers to obtain scaling relations
for compressible MHD turbulence.
We discuss scalings of 
Alfven, fast, and slow modes in both
magnetically dominated (low $\beta$) and gas pressure dominated (high $\beta$)
plasmas.
We also show that the new regime of MHD turbulence
below viscous cutoff 
reported earlier for incompressible flows persists for
compressible turbulence.
Our recent results show that this
leads to density fluctuations.
New understanding of MHD turbulence
is likely to influence
many key astrophysical problems.
}

\keywords{Magnetic fields --- MHD, Diffuse Medium --- ISM:
    Galaxies, Magnetic Fields --- Sun} 

\begin{document}

\maketitle
\section{ISM and MHD Turbulence}
The interstellar medium (ISM) in spiral 
galaxies is crucial for determining the galactic evolution and the
history of star formation. The ISM is turbulent on scales ranging from AUs to
kpc (Armstrong et al 1995; Stanimirovic \& Lazarian 2001; Deshpande et
al 2000), with an embedded magnetic field that influences almost all
of its properties. This turbulence holds the key to many astrophysical
processes (star formation, the phases, spatial distribution, heating
of the ISM, and others).

Present codes can produce simulations that resemble observations
(e.g., Vazquez-Semadeni et al 2000, Vazquez-Semadeni 2000). To what
extent do these results reflect reality? A meaningful numerical
representation of the ISM requires some basic non-dimensional
combinations of the physical parameters of the simulation to be
similar to those of the real ISM. One such parameter is the ``Reynolds number",
$Re$, the ratio of the eddy turnover time of a parcel of gas to the
time required for viscous forces to slow it appreciably. A similar
parameter, the ``magnetic Reynolds number", $Rm$, is the ratio of the
eddy turnover time to magnetic field decay time. The properties of
the flows on all scales depend on $Re$ and $Rm$. Flows with $Re<100$
are laminar; chaotic structures develop gradually as $Re$ increases,
and those with $Re\sim10^3$ are appreciably less chaotic than those
with $Re\sim10^7$. Observed features such as star forming clouds are
very chaotic with $Re>10^8$ and $Rm>10^{16}$. The currently available
3D simulations for a $512^3$ grid can have $Re$ and $Rm$ up to $\sim 6000$
and are limited by their grid sizes. 
It should be kept in mind that while low-resolution
observations show true large-scale features, low-resolution numerics
may produce a completely incorrect physical picture.

How feasible is it, then, to strive to understand the complex
microphysics of astrophysical MHD turbulence? Substantial progress in
this direction is possible by means of ``{\it scaling laws}", or analytical
relations between non-dimensional combinations of physical quantities
that allow a prediction of the motions over a wide range of $Re$.
Even with its limited resolution, numerical simulation is a great tool to {\it
test} scaling laws.

 In spite of its complexity, turbulent
cascade is remarkably self-similar. The physical variables are
proportional to simple powers of the eddy size over a large range of
sizes, leading to scaling laws expressing the dependence of certain
non-dimensional combinations of physical variables on the eddy
size. Robust scaling relations can predict turbulent properties on the
whole range of scales, including those that no large-scale numerical
simulation can hope to resolve. These scaling laws are extremely
important for obtaining insights of processes on the small scales.

By using scaling arguments, Goldreich \& Sridhar (1995) made
ingenious predictions regarding relative motions parallel and
perpendicular to magnetic field {\bf B} for Alfvenic turbulence. 
These relations have been confirmed 
numerically (Cho \& Vishniac 2000; Maron \& Goldreich 2001;
Cho, Lazarian \& Vishniac 2002a; see also Cho, Lazarian \& Vishniac 2002b,
henceforth CLV02 for a review); they are in good
agreement with observed and inferred astrophysical spectra (CLV02). A
remarkable fact revealed in Cho, Lazarian \& Vishniac (2002a) is that
fluid motions perpendicular to {\bf B} are identical to hydrodynamic
motions. This provides an essential physical insight and explains why
in some respects MHD turbulence and hydrodynamic turbulence are
similar, while in other respects they are different.

The Goldreich \& Sridhar (1995, henceforth GS95) model considered 
incompressible MHD,
but the real ISM is highly compressible. This motivated further
studies of the compressible mode scalings.

\begin{figure*}[h!t!]
\includegraphics[width=0.3\textwidth]{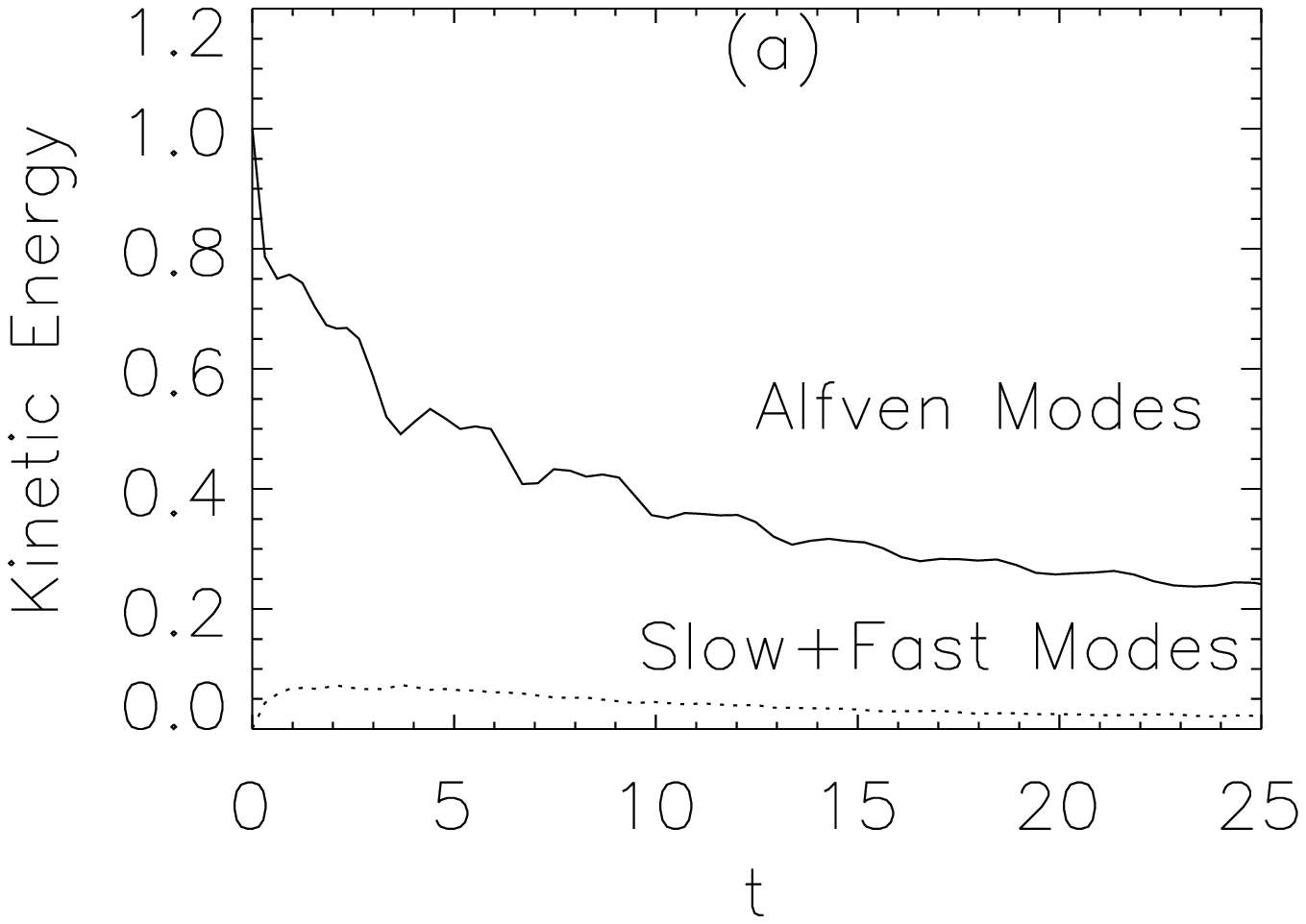}
\hfill
\includegraphics[width=0.3\textwidth]{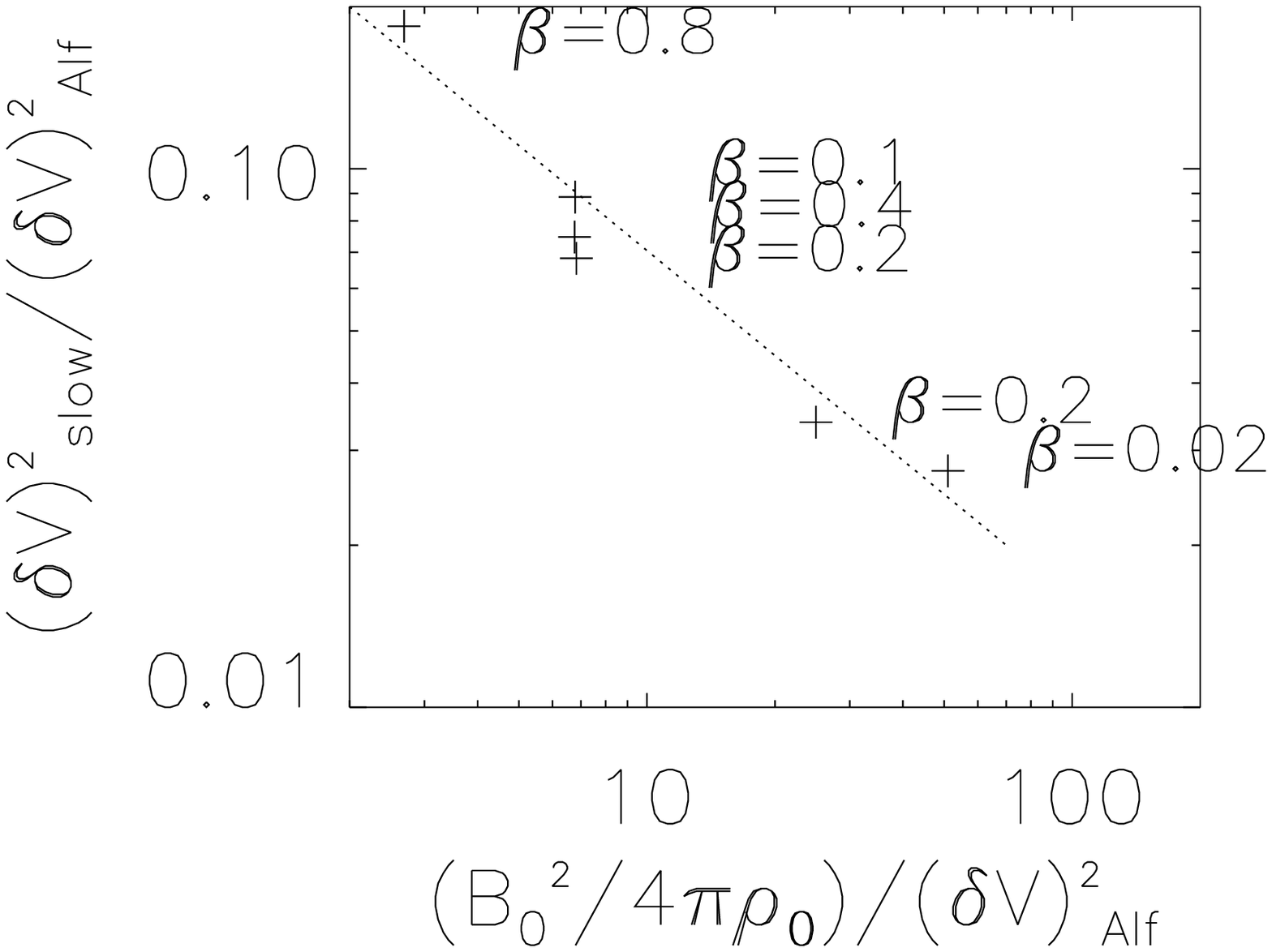}
\hfill
\includegraphics[width=0.3\textwidth]{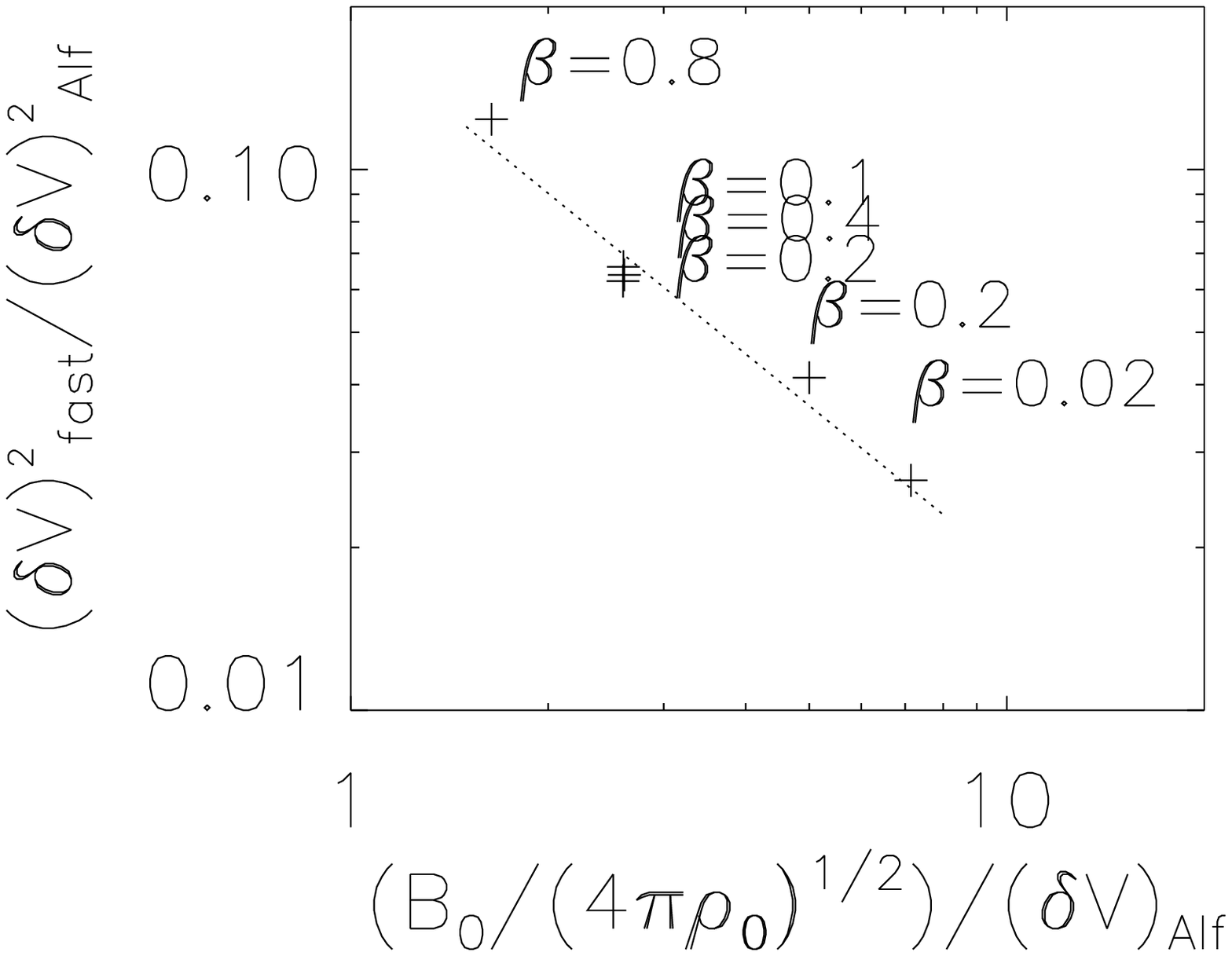}
\caption{ {\it Generation of compressible modes from decaying Alfven Turbulence.}
           Generation is marginal. 
     {\it (Left Panel)} Time evolution. Initially only Alfven modes exist.
     As time goes on, compressible modes are spontaneously generated.
     Initially, $\beta \sim 0.2$ and $B_0/\sqrt{4 \pi \rho_0}=1$.
   {\it (Middle Panel)} Generation of slow modes as a function of $B_0^2$.
   {\it (Right Panel)} Generation of fast modes as a function of $B_0$.
   {}From Cho \& Lazarian (CL02).
       }
\label{fig_generation}
\end{figure*}  

\section{Effects of Compressibility}

Three types of waves exist in compressible magnetized plasma. They are
Alfven, slow and fast waves. Turbulence is a highly non-linear phenomenon
and it has been thought that it may not be productive to talk about
different types of perturbations or modes in compressible media as
everything is messed up by strong interactions. 
Although the statements of this
sort dominate the MHD literature one may question how valid they are.
If, for instance, one particular type of perturbations cascades to small
scales faster than it interacts with the other types, its cascade should
proceed on its own. Within the GS95 model the Alfvenic perturbations
cascade to small scales over their period, while the other 
non-linear interactions
are longer. Therefore we would expect that the GS95 scaling for Alfven modes
should be present in compressible turbulence as well.

Some hints about effects of compressibility can be inferred from Goldreich
\& Sridhar (1995) seminal paper. A more focused theoretical analysis was
done in Lithwick \& Goldreich (2001) paper which deals with properties
of pressure dominated plasma, i.e.  in high $\beta$ regime, where 
$\beta\equiv P_{gas}/P_{mag}\gg 1$. Incompressible regime formally
corresponds to $\beta\rightarrow \infty$ and therefore it is natural
to expect that for $\beta\gg 1$ the turbulence picture proposed in GS95 would
persist. The only actual difference between subsonic turbulence in
high $\beta$ and incompressible regimes is that for finite $\beta$
a new type of motions, fast modes, is present. Those modes for pressure
dominated environments are analogous to sound waves which are marginally   
affected by magnetic field. Therefore it is natural to expect that
they will not distort the GS95 picture of MHD cascade. Lithwick \&
Goldreich (2001) also speculated that for low $\beta$ plasmas the GS95
picture may still be applicable.

A systematic study of the compressibility in low $\beta$ plasmas was
done by Cho \& Lazarian (2002, henceforth CL02). 
First of all, we tested the coupling
of compressible and incompressible modes. If Alfvenic modes produce
a copious amount of compressible modes, the whole picture of independent
Alfvenic turbulence fails. However, the calculations 
(see Fig~\ref{fig_generation}) show
that the amount of energy drained into compressible motions is negligible,
provided that the external magnetic field is sufficiently strong. 

\begin{figure*}[h!t!]
\includegraphics[width=1.0\textwidth]{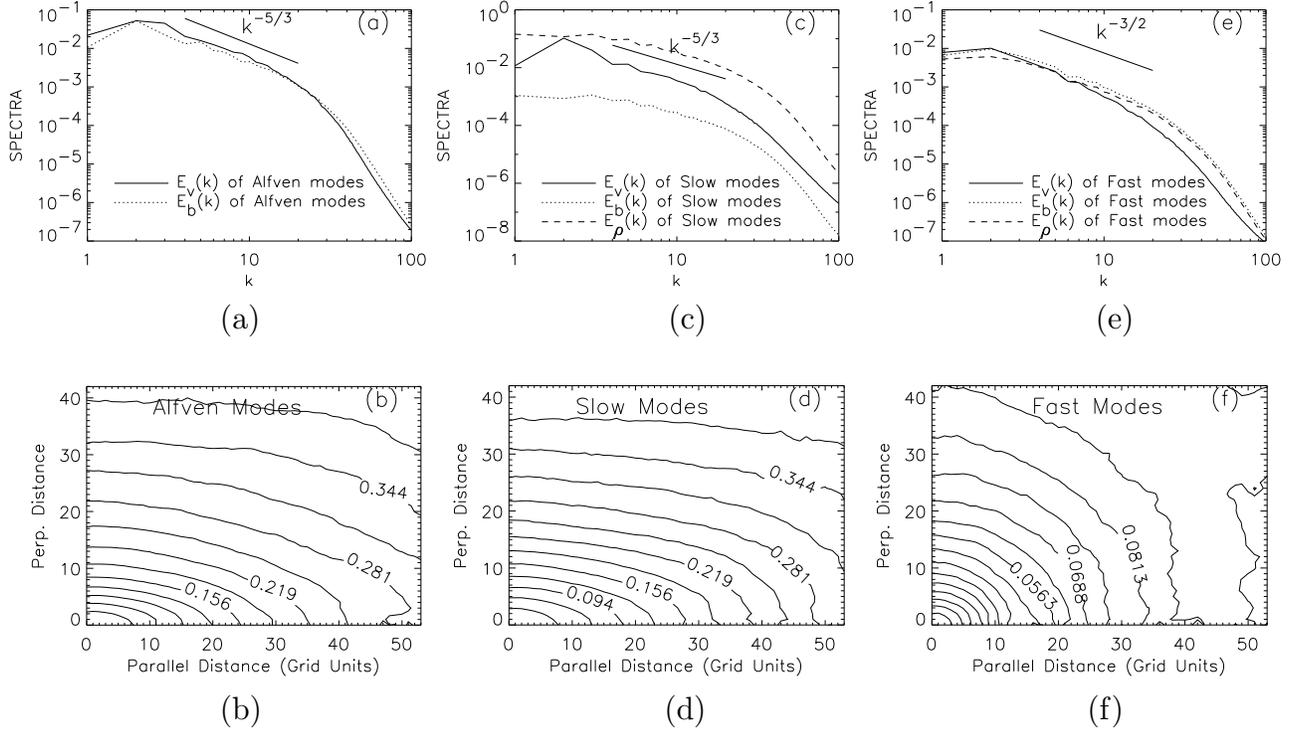}
\caption{ {\it Low $\beta$.}
           Spectra and eddy shapes from driven turbulence with
          $M_s\sim 2.2$, $M_A\sim 0.7$, $\beta\sim 0.2$, 
          and $216^3$ grid points.
         (a) Spectra of Alfv\'en modes follow a Kolmogorov-like power law.
         (b) Eddy shapes 
             (contours of same second-order structure function, $SF_2$)
             for velocity of Alfv\'en modes
             shows anisotropy similar to the GS95
        ($r_{\|}\propto r_{\perp}^{2/3}$ or $k_{\|}\propto k_{\perp}^{2/3}$).
         The structure functions are measured in directions perpendicular or
             parallel to the local mean magnetic field in real space.
             We obtain real-space velocity and magnetic fields 
             by inverse Fourier transform of
             the projected fields.
         (c) Spectra of slow modes also follow a Kolmogorov-like power law.
         (d) Slow mode velocity shows anisotropy similar to the GS95.
             We obtain contours of equal $SF_2$ directly in real space
             without going through the projection method,
             assuming slow mode velocity is nearly parallel to local mean
             magnetic field in low $\beta$ plasmas.
         (e) Spectra of fast modes are compatible with
             the IK spectrum.
         (f) The magnetic $SF_2$ of 
             fast modes shows isotropy.
             We obtain real-space magnetic field 
             by inverse Fourier transform of
             the projected fast magnetic field.
             Fast mode velocity also shows isotropy.
            {}From CL02.
       }
\label{fig_lowbeta}
\end{figure*}  

When the decoupling of the modes was proved, it became meaningful to talk
about separate cascades. A remarkable feature of the GS95 picture is that
turbulent cascade of Alfven waves happens over just one wave
period. Therefore the non-linear interactions with other types of waves
affect the cascade only marginally and the GS95 scaling is expected
to persist in the compressible medium. Moreover, as the Alfven waves
are incompressible, their cascade persists even for supersonic
turbulence. 
In low $\beta$ regime the slow modes are sound-type
perturbations moving along magnetic fields with velocity $a$ 
(or, $a \cos \theta$ in the direction of wave vector ${\bf k}$),
where $a$ ($=\sqrt{\gamma P_{gas}/\rho}$)
is the sound velocity and $\theta$ is the angle between the
wavevector and magnetic field. In magnetically dominated
environments $a\ll V_A$, the gaseous perturbations are essentially static
and the magnetic field mixing motions are expected to mix density
perturbations as if they were passive scalar. As the 
passive scalar shows the same scaling as the velocity field of the inducing
turbulent motions, the slow waves are expected to demonstrate GS95 scalings.
The fast waves in low $\beta$ regime propagate at $V_A$ irrespectively
of the magnetic field direction. Thus the mixing motions induced by
Alfven waves should affect the fast wave
cascade only marginally. The latter cascade should be analogous to the
acoustic wave cascade and be isotropic.

To separate the fast, Alfven and slow modes of MHD turbulence was the
task solved in CL02. Earlier researchers were probably not much motivated
to attempt this, as it was generally believed that the different modes
are messed up in turbulence anyhow. 
To separate the modes, let us consider the displacement vector $\xi$.
For example, $\xi_f({\bf k})$ denotes the direction of displacement for
the fast mode with wave vector ${\bf k}$.
In CL02, we showed that fast, slow, and Alfven displacement vectors 
($\xi_f$, $\xi_s$, and $\xi_A$, respectively) can be
written in terms of ${\bf k}_{\|}$ and ${\bf k}_{\perp}$:
\begin{eqnarray}
   {\bf \xi}_f({\bf k}) \propto 
     \frac{ -1 + \alpha + \sqrt{D} }{ 1+\alpha + \sqrt{D} }
            k_{\|} \hat{\bf k}_{\|} 
     + 
      k_{\perp} \hat{\bf k}_{\perp},    \\
     {\bf \xi}_s({\bf k}) \propto 
            k_{\|} \hat{\bf k}_{\|} 
     + 
     \frac{ 1+\alpha - \sqrt{D} }{ -1 + \alpha - \sqrt{D} }
      k_{\perp} \hat{\bf k}_{\perp},  \\
     {\bf \xi}_A({\bf k}) \propto {\bf k}_{\|} \times {\bf k}_{\perp},
\end{eqnarray}
where ${\bf k}_{\|}$ (or ${\bf k}_{\perp}$) is the wave vector parallel (or
perpendicular) to the mean field ${\bf B}_0$,
$\alpha=a^2/V_A^2=\beta \gamma/2$, and $D=(1+\alpha)^2-4\alpha \cos^2\theta$.
After proper normalization, we get a set of unit bases 
(${\bf \xi}_f({\bf k}), {\bf \xi}_s({\bf k}),{\bf \xi}_A({\bf k})$).
These unit vectors are well-defined and mutually perpendicular except on the
${\bf k}_{\|}$ axis.
We can get fast, slow, or Alfven component velocity by
projecting the Fourier velocity component $\hat{\bf v}({\bf k})$ onto
these unit bases.
See CL02 and Cho \& Lazarian (preprint) for details.

\begin{figure}[t!]
\includegraphics[width=.48\columnwidth]{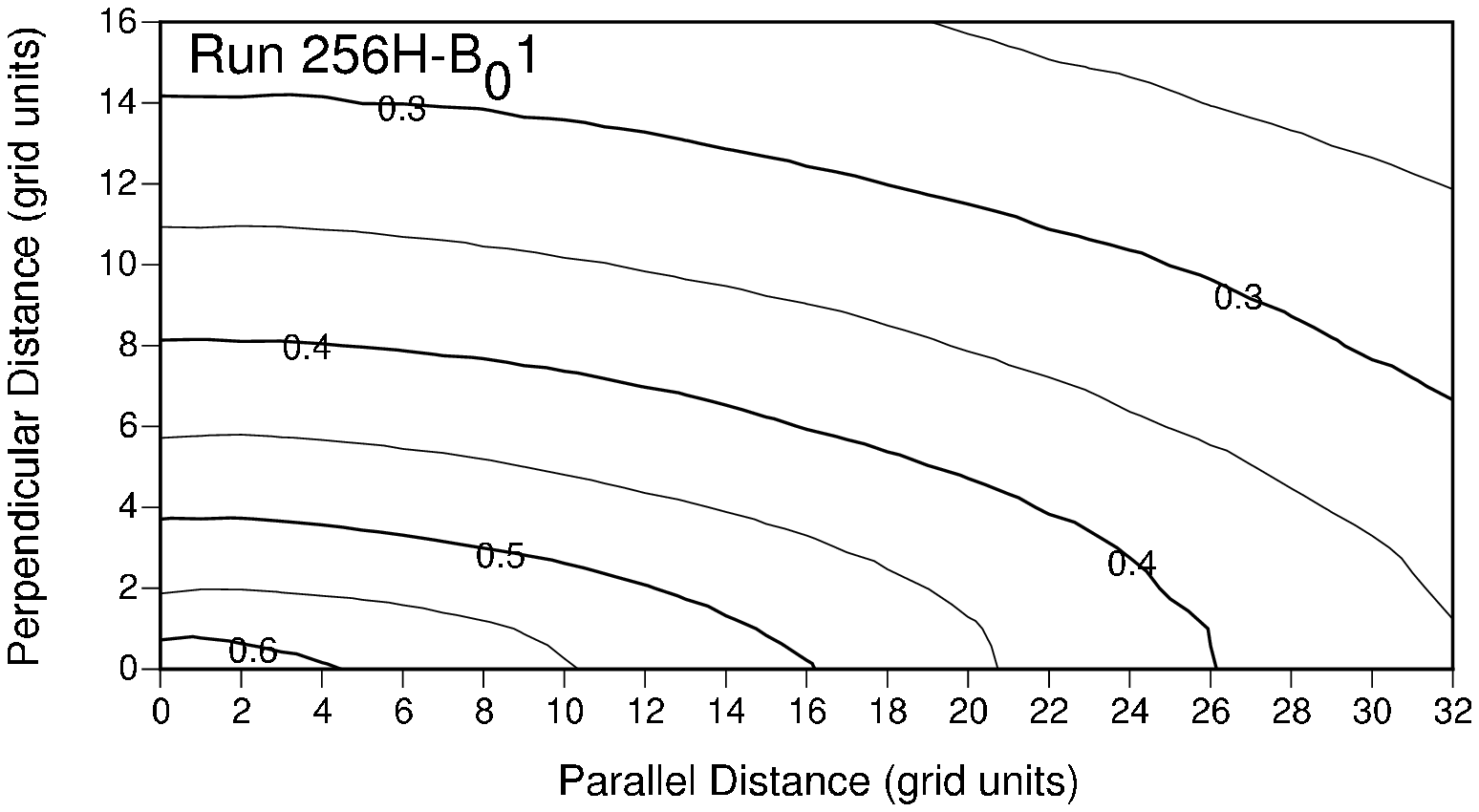}
\hfill
\includegraphics[width=.48\columnwidth]{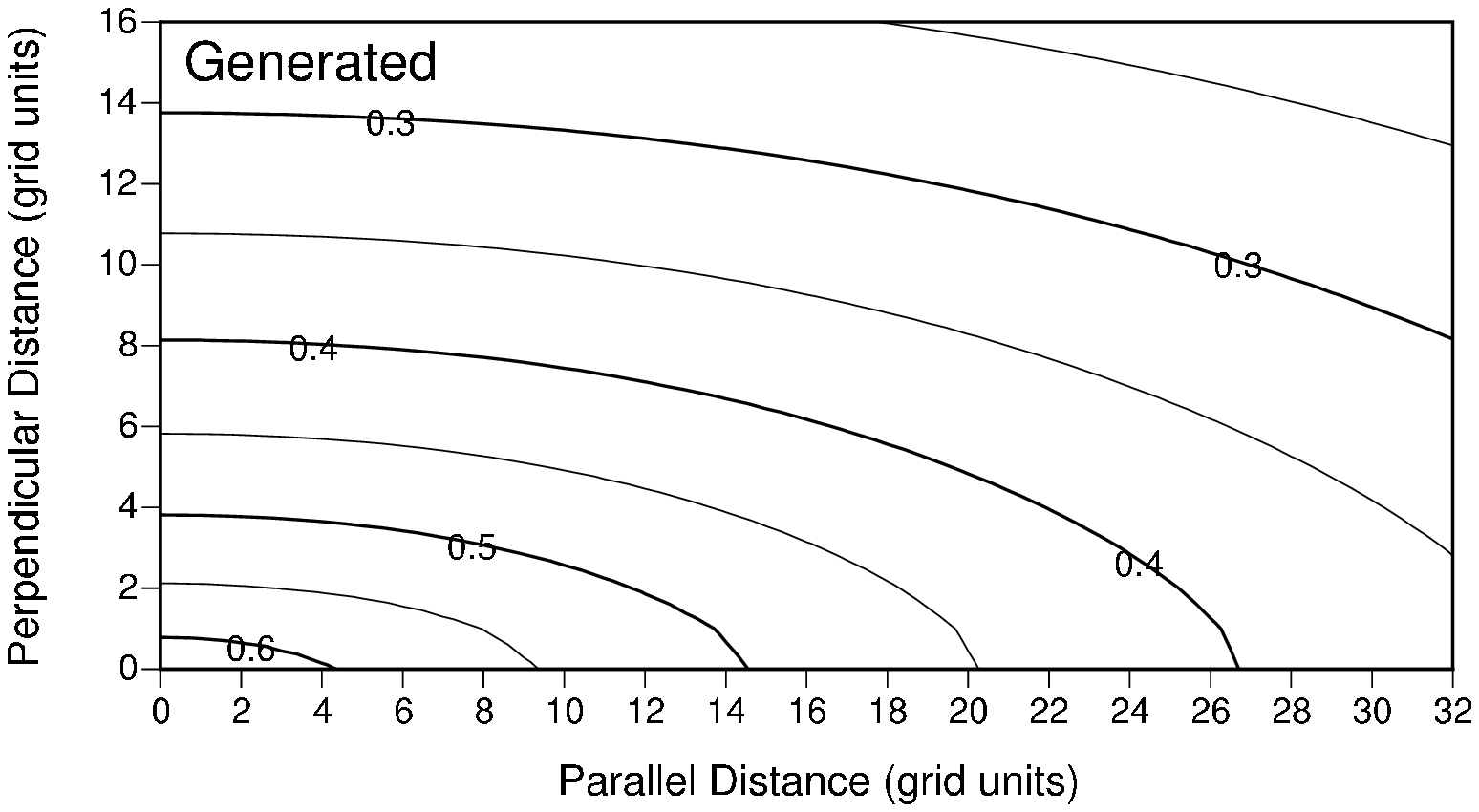}
\caption{
      (a) Iso-contours of equal (velocity) correlation from a simulation.
             The contours represent shape of different size eddies.
             The smaller contours (or, eddies) are more elongated.
      (b) Contours generated from analytical formula.
             {}From Cho, Lazarian, \& Vishniac (2002a).
       }
\label{fig_conto}   
\end{figure}

Results of the  CL02 study are shown in Fig~\ref{fig_lowbeta}. 
The calculations confirm 
the theoretical considerations provided above. Indeed, both the spectra
and anisotropies obtained for Alfven and slow modes look remarkably similar
to the earlier incompressible runs (see Fig~\ref{fig_conto} 
where both the data and
analytical fit for the isocontours of equal correlation are shown).

\begin{figure*}[t]
\includegraphics[width=1.0\textwidth]{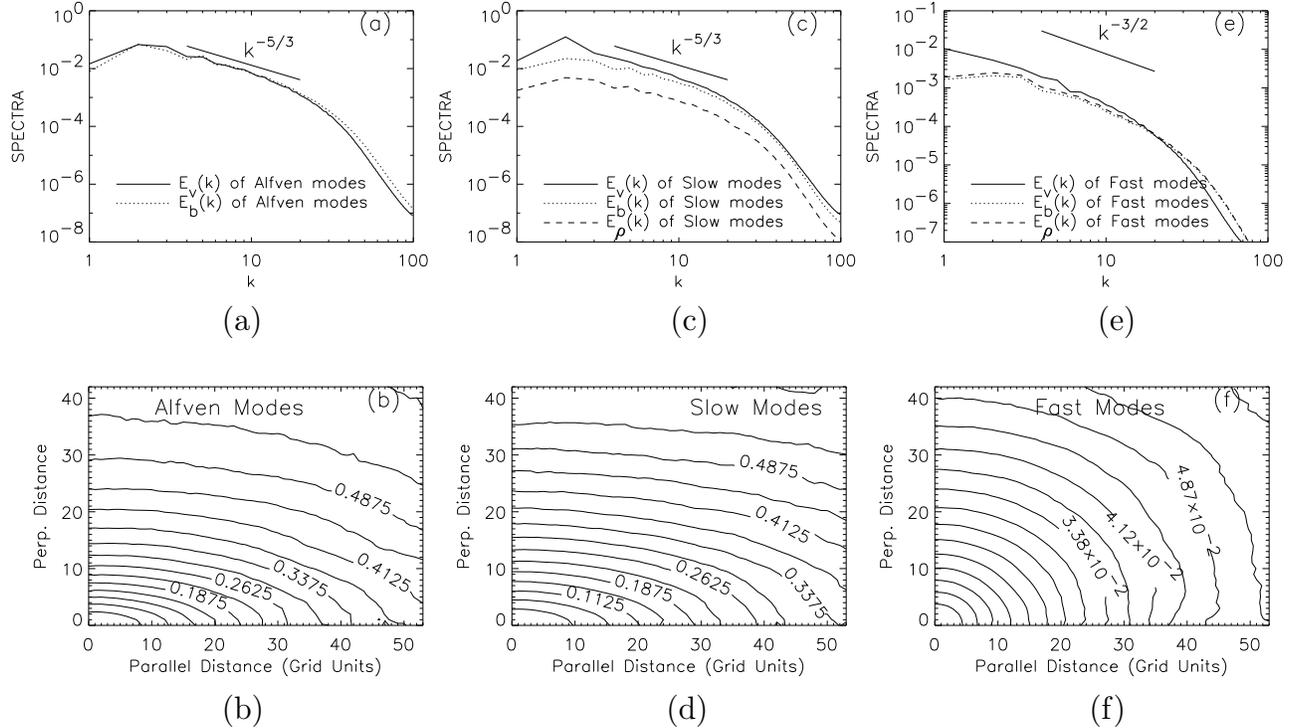}
\caption{ {\it High $\beta$.}
           Spectra and eddy shapes from driven turbulence with
          $M_s\sim 0.5$, $M_A\sim 0.7$, $\beta\sim 4$, 
          and $216^3$ grid points.
          See captions in Fig.~\ref{fig_lowbeta}.
          {}From Cho \& Lazarian (in prep.).
       }
\label{fig_highbeta}
\end{figure*}

Our more recent results for high $\beta$ plasma are shown 
in Fig~\ref{fig_highbeta}. They
confirm theoretical considerations in Lithwick \& Goldreich (2001).
This means that for the first time we have a theoretical insight 
and simple scaling relations working for high $\beta$ and 
low $\beta$ MHD turbulence. What will happen in the regime of moderate
$\beta$?

We would speculate that due to its rapid cascade the Alfven waves should
preserve GS95 scaling. Moreover, the passive nature of the slow modes
in both high and low $\beta$ regimes induce us to think that they would
mimic the spectrum of Alfven modes in the $\beta\sim 1$ case as well.
The decoupling of the fast waves and their isotropy in this regime
are also likely. Thus we may hope that we have a picture of MHD turbulence
in general and the difference between the aforementioned modes 
would stem from the difference in the partition of energy between
magnetic and gas energies in the compressional
modes as $\beta$ varies. For instance, most energy of the slow modes
in low $\beta$ plasma is in gas compression, while in high $\beta$ plasma
the slow modes energy is mostly magnetic.

How important is the strength of the regular magnetic field? In our
computations the magnetic field was taken to be sufficiently strong.
If the magnetic field is weak and strongly entangled by turbulence,
the decoupling of modes is difficult. One may wonder to what
extend our insight above is applicable. We believe that if the 
magnetic field energy $B^2/8\pi$ is much smaller than energy $\rho V_l^2/2$
at a scale $l$, the magnetic field would not change the dynamics of
eddies at this scale and therefore the turbulence would be hydrodynamic.
At smaller scales, however, as the energy of eddies decreases (e.g.
as $l^{2/3}$ for the Kolmogorov cascade) the magnetic field becomes
dynamically important with the implication that the scaling relations
tested for the strong external magnetic field should be applicable.  

\section{Ion-neutral damping: a new regime of turbulence} 

In hydrodynamic turbulence viscosity sets a minimal scale for
motion, with an exponential suppression of motion on smaller
scales.  Below the viscous cutoff the kinetic energy contained in a 
wavenumber band is 
dissipated at that scale, instead of being transferred to smaller scales.
This means the end of the hydrodynamic cascade, but in MHD turbulence
this is not the end of magnetic structure evolution.  For 
viscosity much larger than resistivity,
$\nu\gg\eta$, there will be a broad range of
scales where viscosity is important but resistivity is not.  
On these
scales magnetic field structures will be created 
by the shear from non-damped turbulent motions, which
amounts essentially to the shear from the smallest undamped scales.
The created magnetic structures would evolve through
generating small scale motions.
As a result, we expect
a power-law tail in the energy distribution, rather than an exponential
cutoff.  This  completely new regime
for MHD turbulence was reported in Cho, Lazarian \& Vishniac (2002c).
Further research showed that there is a smooth connection between this
regime and small scale turbulent dynamo in high Prandtl number fluids
(see Schekochihin et al. 2002).

In partially ionized gas
neutrals produce viscous damping of turbulent motions. 
In the Cold Neutral Medium (see Draine \& Lazarian 1999 for a list of
the idealized phases) this produces damping on the scale of a fraction of
a parsec. The magnetic diffusion in those circumstances is
still negligible and exerts an influence only 
at the much smaller scales, $\sim 100km$. 
Therefore, there is a large
range of scales where the physics of the turbulent cascade 
is very different from 
the GS95 picture.

\begin{figure}[t]
\includegraphics[width=.49\columnwidth]{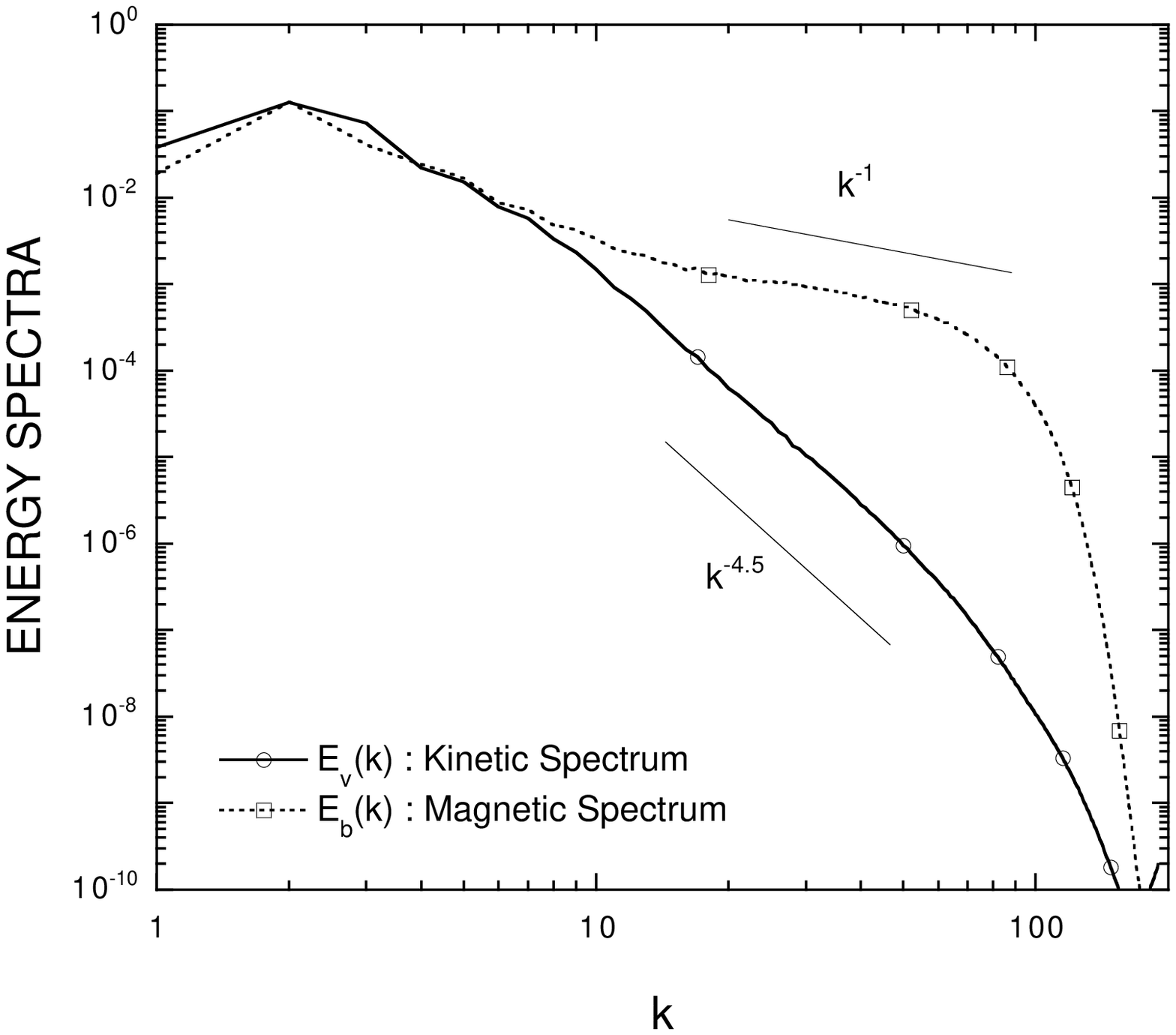}
\hfill
\includegraphics[width=.49\columnwidth]{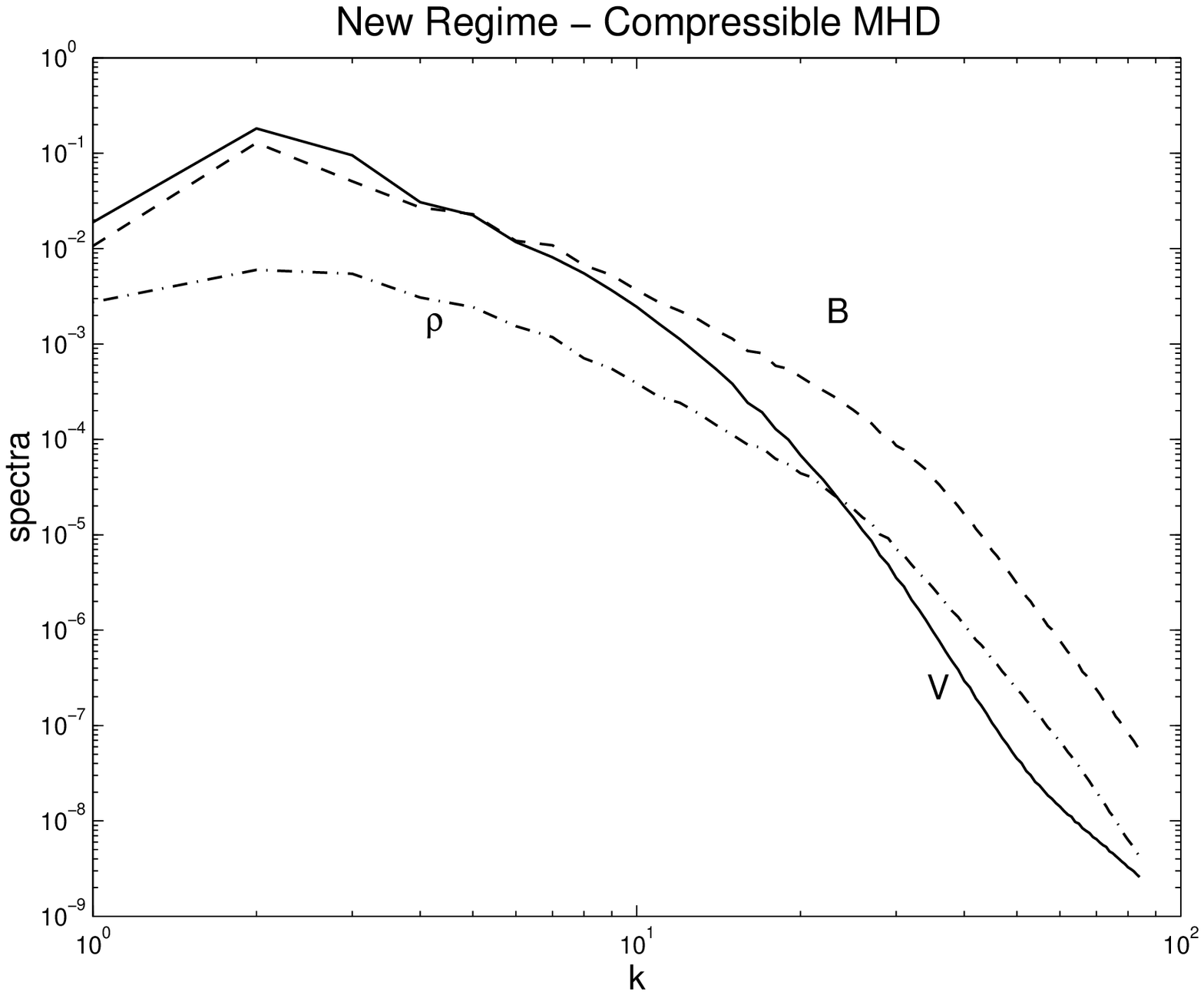}
\caption{ {\it New Regime of MHD turbulence: viscously damped.}
         (a) {\it (Left)} Incompressible case.
             A new inertial range
             emerges below the viscous cut-off at $k\sim 7$.
             From Cho, Lazarian, \& Vishniac (2002c).
         (b) {\it (Right)} Due to numerical reasons 
		the inertial range is truncated when
		we use our compressible code. However, the 
		new regime persists and causes density fluctuations.  
                {}From an upcomming paper.
       }
\label{fig_imb}
\end{figure}

Cho, Lazarian, \& Vishniac (2002c) 
explored this regime numerically with a grid of $384^3$ and a 
physical viscosity for velocity damping. The kinetic Reynolds number was
around 100. The result is presented in  Fig.~\ref{fig_imb}a.

A theoretical model for this new regime and its
consequences for stochastic reconnection (Lazarian \& Vishniac 1999) 
can be found
in Lazarian, Vishniac, \& Cho (2002). 
It explains the spectrum $k^{-1}$ as a cascade of magnetic
energy to small scales under the influence of shear at the 
marginally damped scales. Moreover, our work suggests that the 
magnetic fluctuations protrude to the decoupling scales and
cause the renewal of the MHD cascade there. Earlier work, e.g.
Lithwick \& Goldreich (2001) argued that the turbulent 
cascade survives ion-neutral damping only when a high degree of
ionization is present. In view of the finding a revision of
a few earlier theoretical conclusions is necessary. It should 
be noted that our conclusion about the resumption of turbulence
at small scales is consistent with observations
(Spangler 1991; 1999) that do not show any change of
the observed electron scintillation spectrum at the ambipolar
damping scale. 

We show our results for the compressible fluid in Fig~\ref{fig_imb}b. 
The inertial 
range is much smaller due to numerical reasons, but it is clear that
the new regime of MHD turbulence persists. The magnetic fluctuations,
however, compress the gas and thus cause fluctuations in density.
This is a new (although expected) phenomenon compared to our earlier
incompressible calculations. These density fluctuations may have important
consequences for the small scale structure of the ISM.
We may speculate that they might have some relation to 
the tiny-scale atomic structures (TSAS).
Heiles (1997) introduced the term TSAS
for the mysterious
H~I absorbing structures on scales from thousands to tens of
AU, discovered by Dieter, Welch \& Romney (1976). Analogs are observed
in NaI and CaII (Meyer \& Blades 1996; Faison \& Goss 2001;
      Andrews, Meyer, \& Lauroesch\ 2001) 
and in molecular gas (Marscher, Moore, \& Bania 1993).

Our calculations  
are applicable on scales from the viscous damping 
scale (determined by equating the energy transfer rate with the 
viscous damping rate; $\sim0.1$ pc in the Warm Neutral Medium with $n$ 
= 0.4 cm$^{-3}$, $T$= 6000 K) to the ion-neutral decoupling scale (the 
scale at which viscous drag on ions becomes comparable to the neutral 
drag; $\ll 0.1$ pc). Below the viscous scale the fluctuations of 
magnetic field obey the damped regime shown in Figure \ref{fig_imb}b and 
produce 
density fluctuations. For typical Cold Neutral Medium gas, the scale of 
neutral-ion decoupling decreases to $\sim70$ AU, and is less for denser 
gas. TSAS may be created by strongly nonlinear MHD turbulence!

\section{Discussion}

In this paper we have discussed the new outlook
onto compressible MHD turbulence. Contrary to common beliefs the compressible
MHD turbulence does not present a complete mess, but demonstrates
nice scaling relations for its modes. A peculiar feature is that
those relations should be studied locally, i.e. in the frame related
to the local magnetic field. However, such a system of reference is
natural for many phenomena, e.g. for cosmic ray propagation. 
Recent application of the scalings obtained for compressible turbulence
have shown that fundamental revisions are necessary for the field of
high energy astrophysics. For instance, Yan \& Lazarian (2002) demonstrated
that fast modes dominate cosmic ray scattering even in
spite of the fact that they are subjected to collisional and collisionless
damping. This entails consequences for models of cosmic ray propagation,
acceleration, elemental abundances etc.

Advances in understanding of MHD turbulence have very broad astrophysical
implications. The fields affected span from accretion disks and stars to the
ISM and the intergalactic medium in clusters. Turbulence is known to hold the
key to many astrophysical processes. It was considered too messy by
many researchers who consciously or subconsciously 
tried to avoid dealing with it. 
Others, more brave types, used Kolmogorov
scalings for compressible strongly magnetized gas, although they
did understand that those relations could not be true. 
\adjustfinalcols
Recent research
in the field provides the scaling relations and insights that will
contribute to many areas of research.   


\acknowledgments

AL and JC acknowledge support by  the NSF grant AST0125544.

\end{document}